\documentclass[psamsfonts]{amsart}

\usepackage{natbib}
\numberwithin{equation}{section}

\pdfoutput=1
\usepackage{graphics}
\usepackage[dvipdf]{graphicx}
\usepackage{epstopdf}
\usepackage{epsfig}
\usepackage{url}

\usepackage{natbib}

\usepackage{epsfig}

\usepackage{epsfig}

\usepackage{amssymb}

\usepackage{pifont}

\usepackage{supertabular}
\usepackage{array}
\usepackage{picture}
\usepackage{arydshln}

\usepackage{tikz}
\usetikzlibrary{snakes}

\begin{document}

\title[ICEQUAKES AND PRECURSORS]{Icequakes coupled with surface displacements for predicting glacier
 break-off}

%

\author{J. Faillettaz}
\address{J. Faillettaz\\VAW, ETH Zurich, Laboratory of Hydraulics, Hydrology and Glaciology\\Switzerland} 
\email{faillettaz@vaw.baug.ethz.ch}
\url{http://www.glaciology.ethz.ch}
\author{ D. Sornette}
\address{DD. Sornette\\epartment of Management, Technology and Economics, ETH Z\"urich}
\address{Department of Earth Sciences, ETH Z\"urich}
\address{Institute of Geophysics and Planetary Physics, UCLA}
\curraddr{Department of Management, Technology and Economics,\\ ETH Z\"urich\\Switzerland}
\email{ dsornette@ethz.ch}
\author{M. Funk}
\address{M. Funk\\VAW, ETH Zurich, Laboratory of Hydraulics, Hydrology and Glaciology\\Switzerland}
\email{funk@vaw.baug.ethz.ch}

\begin{abstract}

A hanging glacier at the east face of Weisshorn (Switzerland) broke off in 2005. We were
able to monitor and measure surface motion and icequake activity for 25 days up to
three days prior to the break-off. The analysis of
seismic waves generated by the glacier during the rupture maturation process revealed
four types of precursory signals of the imminent catastrophic rupture: 
(i) an increase in seismic activity within the glacier, (ii) a
decrease in the waiting time between two successive icequakes, (iii)
a change in the size-frequency distribution of icequake energy, and (iv) a modification
in the structure of the waiting time distributions between two successive icequakes. 
Morevover, it was possible to demonstrate the existence of a correlation between the seismic activity and the
log-periodic oscillations of the surface velocities superimposed on the global
acceleration of the glacier during the rupture maturation.
Analysis of the seismic activity 
led us to the identification of two
regimes: a stable phase with diffuse damage, and an unstable and dangerous phase
characterized by a hierarchical cascade of rupture instabilities where large
icequakes are triggered.
\end{abstract}

\maketitle

\section{Introduction}

The fracturing of brittle heterogeneous material has often been studied at the
laboratory scale using acoustic emission measurements (see for instance
\citet{Johansen&Sornette2000, Nechad&al2005a} for recent observations interpreted
using concepts relevant to the present study).
These studies reported an acceleration of brittle damage before failure.
Acoustic emission tools have already been used at meso-scale to find precursors to
natural gravity-driven instabilities such as cliff collapse
\citep{Amitrano&al2005} or slope instabilities
\citep{Dixon&Spriggs2007, Kolesnikov&al2003,Dixon&al2003}.
The present paper focuses on the acoustic emissions generated by an unstable glacier. 
To our knowledge, this is the first attempt to use these acoustic emissions
to predict the catastrophic break-off of a glacier.

Ice mass break-off is a natural gravity-driven instability as found in the
case of a landslide, rockfalls or
mountain collapse. 
Such glacier break-off represents a considerable risk to mountain communities
and transit facilities situated below, especially in winter, as an ice avalanche may drag snow in
its train. 
In certain cases, an accurate prediction of this natural phenomenon is necessary in order to
prevent such dangerous events.
The first attempt to predict such break-offs was conducted in 1973 by
\citet{Flotron1977} and
\citet{Roethlisberger1981a} on the Weisshorn hanging glacier. 
This latter author measured the surface velocity of the unstable glacier and proposed an empirical function
to fit the increasing surface velocities before break-off. 
This function describes an acceleration of the surface displacement following
a power law up to infinity at a finite time $t_c$. 
Obviously, the real break-off will necessarily occur before $t_c$, but the
method gives a good description of the surface velocity evolution until
rupture.
Recently, following \citet{Luethi2003} and 
\citet{Pralong&al2005}, \citet{Faillettaz&al2008} showed evidence of an oscillatory
behaviour superimposed on the general acceleration which enables a more
accurate determination of the time of rupture.
\citet{Faillettaz&al2008} showed also an increase in icequake activity before the
break-off. The aim of this paper is to present (i) a careful analysis of these seismic
measurements, (ii) our conclusions in terms of rupture processes, and (iii) perspectives
for forecasting.

Several studies have shown that glaciers can generate seismic signals called
icequakes.
Previous studies have identified at least five characteristic
seismic waveforms associated with five different icequake event types. These include: 1)  surface crevassing (high frequency, short duration,
impulsive onsets, \citet{Neave&Savage1970,Deichmann&al2000,Walter&al2008}), 2) calving events (Low
frequency, long duration, non-impulsive onsets, surface waves, \citet{ONeel&al2007,Neetles&al2008}), 3) basal sliding (low frequency,
short duration, no surface waves, \citet{Weaver&Malone1979}) 4) iceberg interaction (low frequency, long duration,
multiple harmonic frequencies, \citet{MacAyeal&al2008}), and 5) hydraulic
transients in glacial water channels (low
frequency, emergent onset, absence of distinct S wave, \citet{StLawrence&Qamar1979}).

In this study, the focus is on seismic activity generated by a cold hanging
glacier before its break-off. The crucial features of this type of glacier are as follows:
(i) there is no sliding at the bedrock, and (ii) the glacier is entirely cold and there is no water within the ice. 
Precursory seismic signals were detected, and
a change in the behavior occurred two weeks before the global rupture. 

\section{Weisshorn glacier and the history of events}
\begin{figure}[t]
\vspace*{2mm}
\begin{center}
\includegraphics[width=0.7\textwidth]{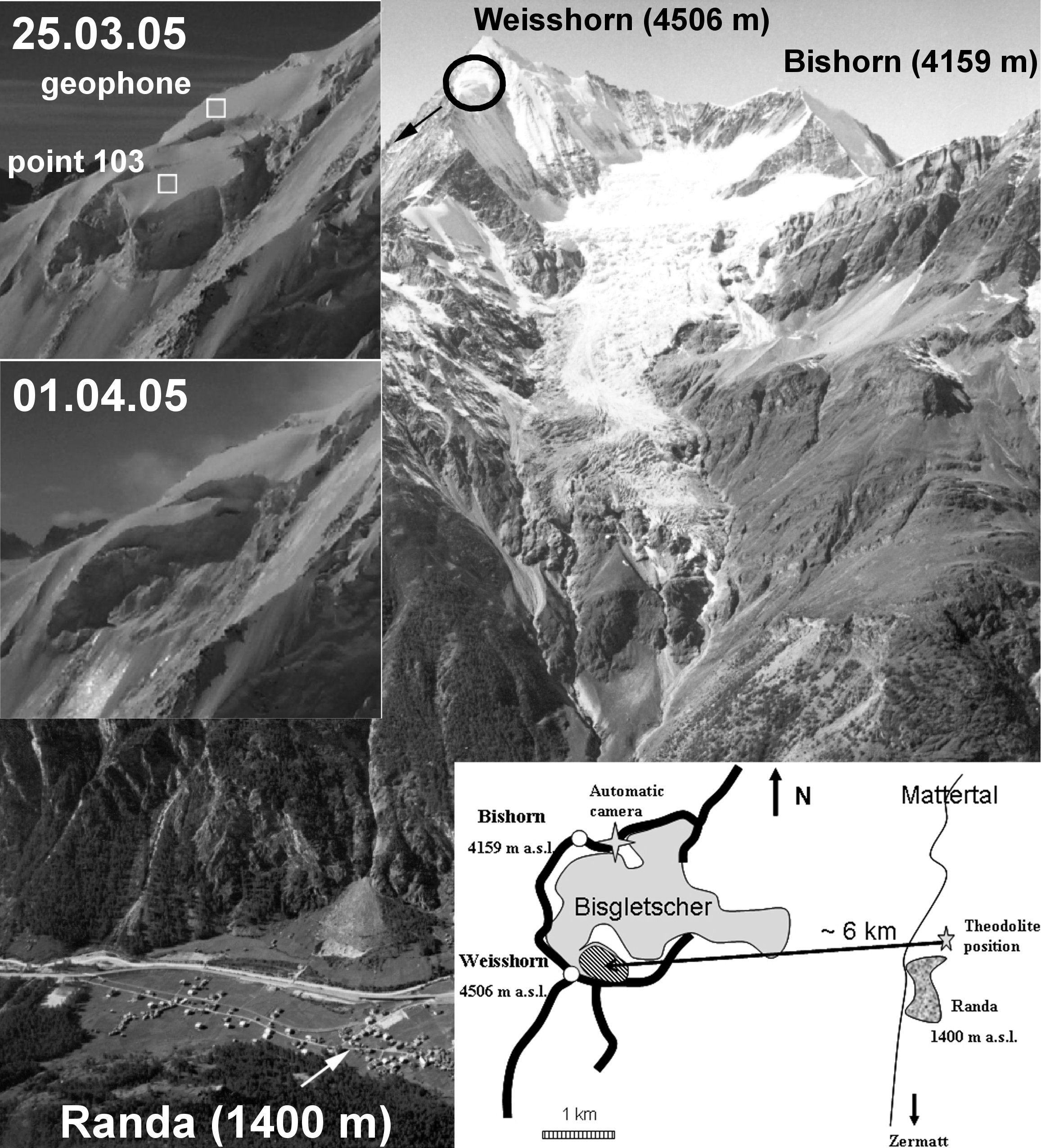}
\end{center}
\caption{\label{general}
The east face of Weisshorn with the hanging glacier. The
village of Randa and transit routes are visible in the valley. The ellipse
indicates the location of the hanging glacier. The left insets shows a closer
frontal view of the hanging glacier on March, 25th 2005 before the second break-off
(upper), and on April 1st, 2005 after the break-off (lower), including the positions
of the geophone and stake 103 used for
displacement measurements.
The bottom right inset gives a general schematic view of the Weisshorn hanging glacier (dashed zone), and
the monitoring setting (theodolite and automatic camera). Thick black lines
indicate the mountain ridges, and the thin line represents the bottom of the valley.}
\end{figure}

The northeast face of the Weisshorn (Valais, Switzerland) is covered with
unbalanced cold ramp glaciers \citep[i.e., the snow accumulation is, for the most part, compensated by
break-off,][]{Pralong&Funk2006}, located between 4500 m and
3800 m a.s.l., on a steep slope of 45 to 50 degrees. 
In winter, snow
avalanches triggered by icefalls pose a recurrent threat to the 400
inhabitants of the village of
Randa located some 2500 m below the glacier, and to transit routes to
Zermatt (see Fig.~\ref{general}) \citep{Raymond&al2003}. In a compilation of
historical records, \citet{Raymond&al2003} showed that, despite no seasonal
pattern in the events, Randa was damaged
repeatedly during the last centuries, always in winter.  Among the 19 events recorded since
1636, three caused a total of 51 fatalities, and six damaged the village of
Randa.
The Weisshorn hanging glacier broke off five times in the last 35
years (1973, 1980, 1986, 1999 and 2005, see \citet{Raymond&al2003}); two of these
events (in 1973 and 2005) were monitored in detail \citep{Flotron1977,Faillettaz&al2008}.

\begin{figure}[t]
\vspace*{2mm}
\begin{center}
\includegraphics[width=0.7\textwidth]{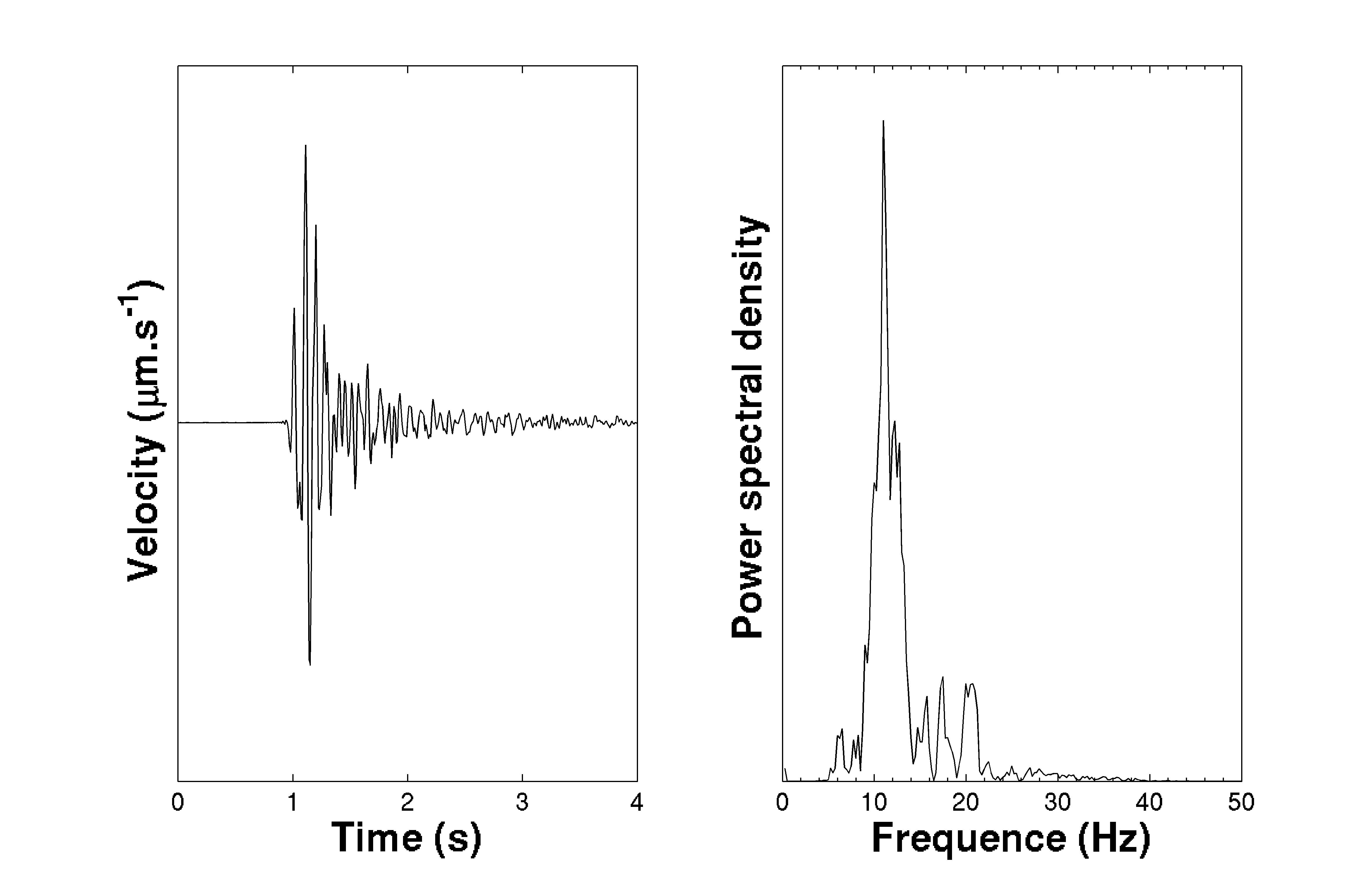}
\end{center}
\caption{\label{spectre}
Unfiltered velocity seismogram of a typical
event (maximum amplitude 2.5 $\rm \mu ms^{-1}$) and its corresponding
normalized power spectrum density (right).}
\end{figure}

The total volume of the unstable ice mass was estimated at $0.5 \times 10^6
~\rm m^3$ by means of photogrammetry \citep{Faillettaz&al2008}.  
Because of the dangerous situation for the village of Randa, a
monitoring system was installed to alert the population of an impending
break-off.

\section{Methods}
\subsection{Instrumentation}

An automatic camera (installed in  September 2003 on the Bishorn, see Fig.~\ref
{general}) provided a detailed movie of the destabilization of the
glacier. A first break-off occurred on March 24, 2005 (after
26.5 days of monitoring). Its estimated volume amounted to $120,000\;~\rm m^3$
(comparable to the 1973 break-off with $160,000\; ~\rm m^3$). On March 31,
2005, a second rupture occurred, during which the major part of the glacier
broke off (after 33.5~days of monitoring). The volume of this second ice avalanche
was estimated at $400,000\; ~\rm m^3$. 


A single geophone (Lennartz LE-3Dlite Mkll, 3 orthogonal sensors, with eigenfrequency of 1 Hz) was installed in firn
30~cm below the surface near the upper crevasse (Fig.~\ref{general}), 
in order to record icequake activity before the final rupture. This signal is
assumed to describe the crack (or damage) evolution within the ice mass during the failure
process. 
A Taurus portable seismograph (Nanometrics inc.) was used to record the seismic
activity of the glacier prior to its rupture with a sampling rate of 100 Hz. 
Unfortunately, the recorder failed on March 21, before the first break-off event, because of
battery problems.
A first seismic analysis of these measurements was presented in \citet{Faillettaz&al2008}.

Concurrently to the seismic measurements, we performed a careful analysis
of the surface displacements of the glacier (see Fig. \ref{timeline}) . 
The measurement equipment consisted of a total station (Leica theodolite
TM1800 combined with the DI3000S Distometer) installed at a fixed position
above Randa on the other side of the valley, and seven reflectors mounted on
stakes drilled into the unstable ice mass. A reference reflector was installed on a rock for the correction of the
measurements, because of broad variations in meteorological conditions. This fully
autonomous apparatus performed the measurements every two hours. The motion of
the reflectors (see Fig.~\ref{general}) could be monitored only when the
visibility conditions were good enough.


\subsection{Analytical methods}

\paragraph{{\bf Icequake detection}}

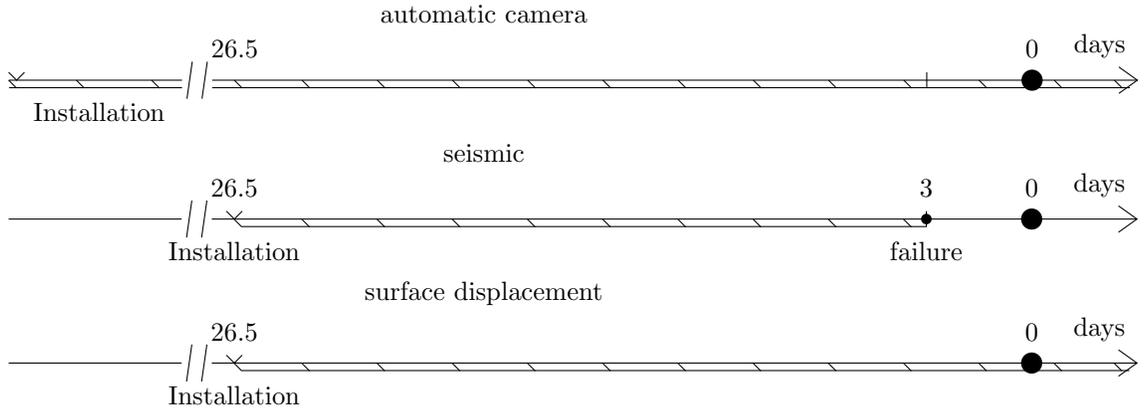
\begin{figure*}[htbp]

\centering automatic camera

	\begin{tikzpicture}
		\draw [-] (0cm, 0cm) -- (2.3cm, 0cm);
\draw [-] (2.7cm, 0cm) -- (15cm, 0cm);
		\draw [-] (0cm,-0.1cm) -- (2.3cm, -0.1cm);
\draw [-] (2.7cm, -0.1cm) -- (14.9cm, -0.1cm);
	\foreach \x in {2.4,2.6} \draw (\x cm-1pt, -7pt) -- (\x cm+1pt,  7pt);
		\foreach \x in {12.2,13.6} \draw (\x cm, 3pt) -- (\x cm, - 3pt);
		\draw (0.1cm - 3pt, 0cm + 3pt) -- (0.1cm, 0cm) -- (0.1cm + 3pt, 0cm
               + 3pt);
\draw (15cm-7pt,0cm+5pt) -- (15cm,0cm) -- (15cm-7pt,0cm-5pt);
\foreach \x in {0.1,1,2,3.1,4,5,6,7,8,9,10,11,12,13,14,14.8} \draw (\x cm - 3pt,0) -- (\x cm, - 3pt);
		\draw (3cm, 0cm) node[above=5pt] {26.5};
\draw (1.2cm, 0cm) node[below=5pt] {Installation};
		\draw (13.6cm, 0cm) node[above=5pt] {0};
\draw (14.5cm, 0cm) node[above=5pt] {days};
		\fill (13.6cm, 0cm) circle (4pt);
	\end{tikzpicture}

\centering seismic

	\begin{tikzpicture}
\draw [-] (0cm, 0cm) -- (2.3cm, 0cm);
\draw [-] (2.7cm, 0cm) -- (15cm, 0cm);
	\draw (15cm-7pt,0cm+5pt) -- (15cm,0cm) -- (15cm-7pt,0cm-5pt);
\draw [-] (3.1cm, -0.1cm) -- (12.2cm, -0.1cm);
	\foreach \x in {2.4,2.6} \draw (\x cm-1pt, -7pt) -- (\x cm+1pt,  7pt);
		\foreach \x in {12.2,13.6} \draw (\x cm, 3pt) -- (\x cm, - 3pt);
		\draw (3cm - 3pt, 0cm + 3pt) -- (3cm, 0cm) -- (3cm + 3pt, 0cm
               + 3pt);
\foreach \x in {3.1,4,5,6,7,8,9,10,11,12} \draw (\x cm-3pt,0) -- (\x cm, - 3pt);
		\draw (3cm, 0cm) node[above=5pt] {26.5};
\draw (3cm, 0cm) node[below=5pt] {Installation};
		\draw (12.2cm, 0cm) node[above=5pt] {3};
\draw (12.2cm, 0cm) node[below=5pt] {failure};
		\draw (13.6cm, 0cm) node[above=5pt] {0};
\draw (14.5cm, 0cm) node[above=5pt] {days};
		\fill (12.2cm, 0cm) circle (2pt);
\fill (13.6, 0cm) circle (4pt);
	\end{tikzpicture}

\centering surface displacement

	\begin{tikzpicture}
\draw [-] (0cm, 0cm) -- (2.3cm, 0cm);
\draw [-] (2.7cm, 0cm) -- (15cm, 0cm);
	\draw (15cm-7pt,0cm+5pt) -- (15cm,0cm) -- (15cm-7pt,0cm-5pt);
\draw [-] (3.1cm, -0.1cm) -- (14.9cm, -0.1cm);
	\foreach \x in {2.4,2.6} \draw (\x cm-1pt, -7pt) -- (\x cm+1pt,  7pt);	
		\foreach \x in {13.6} \draw (\x cm, 3pt) -- (\x cm, - 3pt);
		\draw (3cm - 3pt, 0cm + 3pt) -- (3cm, 0cm) -- (3cm + 3pt, 0cm
               + 3pt);
\foreach \x in {3.1,4,5,6,7,8,9,10,11,12,13,14,14.8} \draw (\x cm-3pt,0) -- (\x cm, - 3pt);
		\draw (3cm, 0cm) node[above=5pt] {26.5};
\draw (3cm, 0cm) node[below=5pt] {Installation};
		\draw (13.6cm, 0cm) node[above=5pt] {0};
\draw (14.5cm, 0cm) node[above=5pt] {days};
\fill (13.6, 0cm) circle (4pt);
	\end{tikzpicture}
	\caption{\label{timeline}Timeline of  monitoring. The origin of time $0$ corresponds
	to the occurrence of the first break-off on March 24, 2005 (after 26.5 days of monitoring)
	with an estimated volume of about $120,000 ~ \rm m^3$. }
\end{figure*}

We identified seismic events both visually and by using an automatic earthquake
detection method 
based on the ratio of the root-mean-square between the short-term average (STA) window
and the long-term average (LTA) window. 
The detection of events was performed in the following way. First, we evaluated
the root-mean-square (rms) of two concurrent time windows. The rms values over
the previous 800 ms long-term average (LTA) window and the previous 80 ms
short-term average (STA) window were calculated and compared. When the ratio
$\gamma=\rm STA/LTA$ exceeded a given threshold (taken here equal to 3), an event was
detected and retained \citep{Allen1978,Walter&al2008}.
Both of these catalogues (with visual and automated detection) are compatible with each other and give a total number of 1731
icequakes during the monitoring period (see Fig. \ref{timeline}).

\paragraph{{\bf Icequake characterization}}

For a much deeper analysis and to allow a comparison of the detected icequakes,  
their sizes were first evaluated.
Seismic event sizes were estimated based on their signal energies as defined for a
digitalized signal by \citet{Amitrano&al2005}:
\begin{equation}
E=\sum A^2\delta t \;,
\end{equation}
where $A$ is the signal amplitude and $\delta t$ is the sampling period.
We made a manual selection of the beginning and the end of each of the 1731 signals and performed the
discrete summation for the evaluated duration of each event.


Finally, these methods enabled a catalogue of events to be obtained,
containing time of occurrence and the respective energy for each detected icequake.
It was then possible to analyse this catalogue with statistical tools and methods developed for
earthquake study.

\section{Results}

\subsection{Signal characteristics}

The data show a high seismic emissivity from the hanging glacier during the time span of our observations.
In the case of the Weisshorn hanging glacier, seismic events with short and impulsive
signals and similar spectra were observed (see Fig. \ref{spectre}), with dominant
power contained in the 10-30 Hz frequency band. This observation is consistent with
previous results \citep{Neave&Savage1970,Deichmann&al2000,Roux&al2008,ONeel&al2007}. 

This result is not surprising, as no serac falls could be observed during the time
span of our observations (based on daily photographs). 
Since the sensor was very close to the sources, attenuation was low.
The proximity of the source (less than 300 meters) gave rise to difficulties
in distinguishing P and S waves. 

As the geophone was situated above the upper crevasse separating the active from
the stable zone, compressive seismic waves
(primary wave) were perturbed by the discontinuity of the material and
therefore were less likely to be observed.

\begin{figure}
\vspace*{2mm}
\begin{center}
\includegraphics[width=0.7\textwidth]{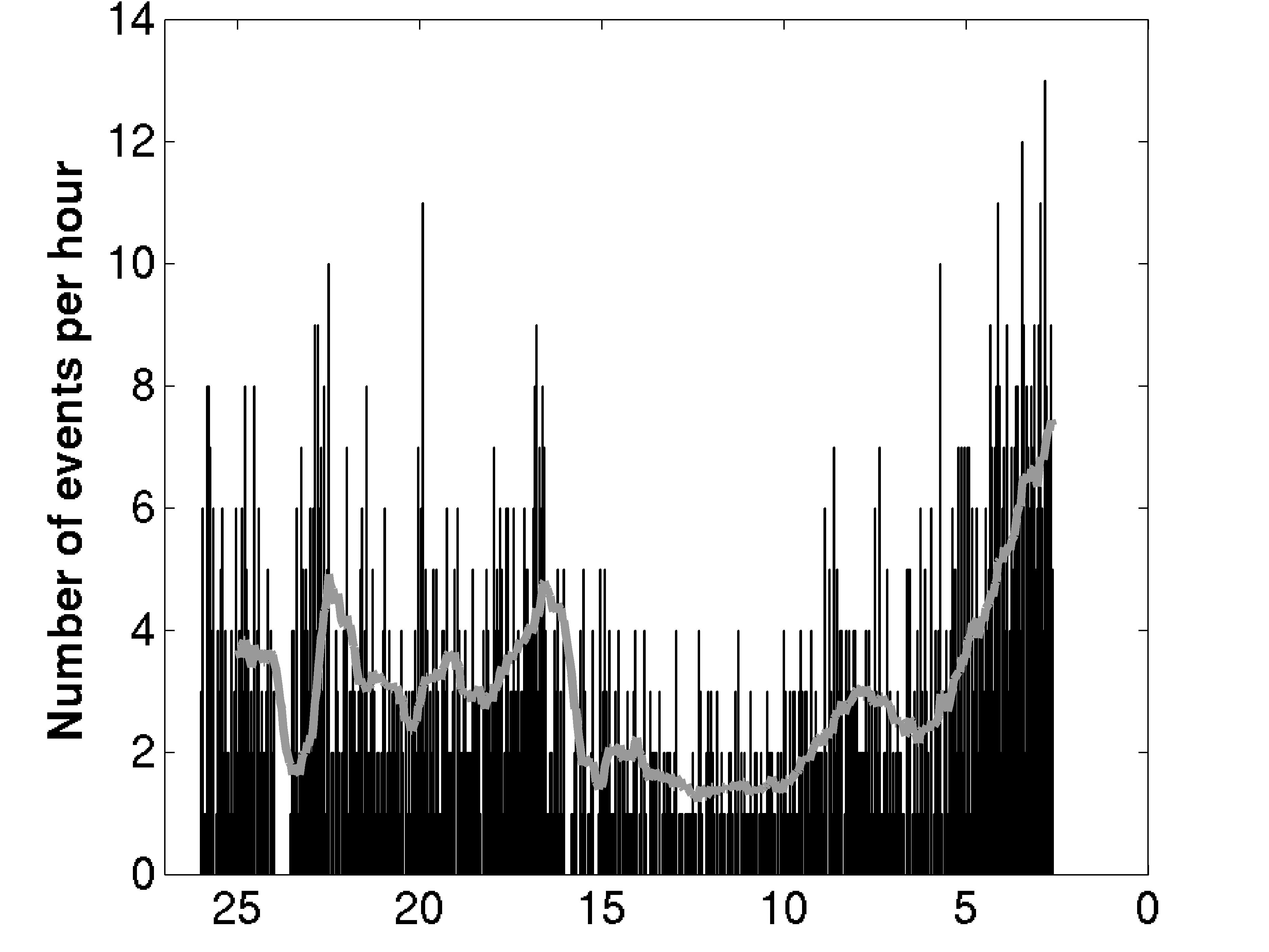}
\end{center}
\caption{\label{fig2} Number of detected icequakes per hour (bars in black) as
a function of time shown on the abscissa. The smoothed number of icequakes per hour, shown
as the light grey line, was obtained by averaging in a sliding window of 24 hours.
}
\end{figure}

Figure \ref{fig2} shows the number of detected events per hour during the 25-day period of the ice chunk destabilization.
An acceleration of the seismic activity was detected 
one week before the first break-off
(i.e., two weeks before the main
break-off).

\subsection{Size-frequency distribution of icequake energies}
\label{energy}


\begin{figure*}
\vspace*{2mm}
\begin{center}
\includegraphics[width=\textwidth]{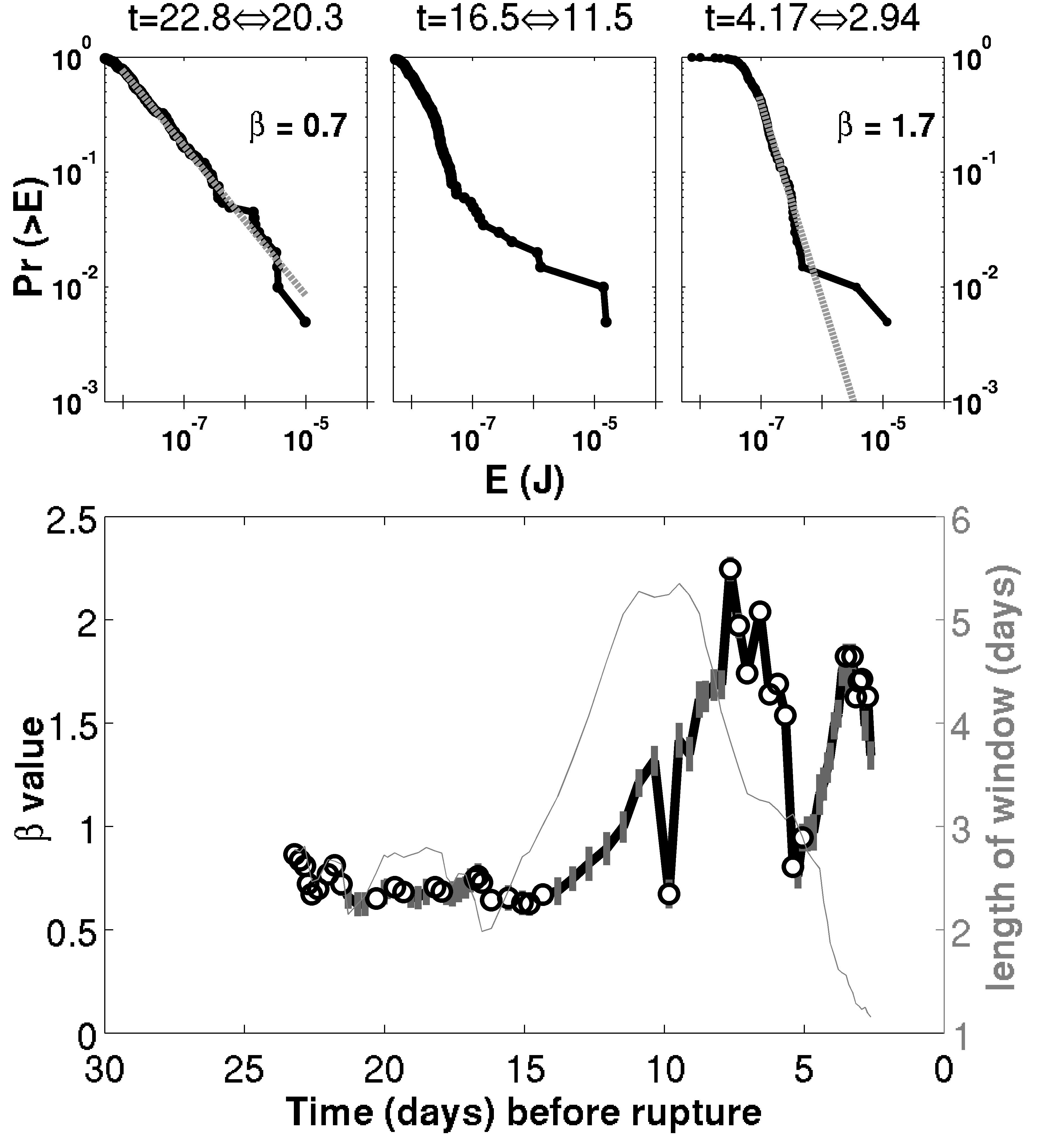}
\end{center}
\caption{\label{fig3}  
The three plots at the top show the complementary cumulative size-frequency distribution (CSFD) ($Pr(>E)$) of
icequake energies (E) obtained in three windows of 200 events each, ending at the
time indicated in the panels. 
The lower plot
shows the evolution of the exponent $\beta$ of the power law fitting the CSFD
obtained in running windows of 200 events. The exponent $\beta$
has been estimated using the Maximum Likelihood method (see text). 
The thin line also gives the duration of the sliding window of 200 events, corresponding
to the scale on the right. The vertical lines indicate the errors given by the
Maximum Likelihood method. 
Empty symbols indicate those fits whose p-value is greater than 0.2,
i.e., for which the power-law behavior is plausible (see text).
}
\end{figure*}

The complementary cumulative size-frequency (also called ``survival'')  distribution (CSFD) of
the icequake energy preceding the break-off was then determined.
The complementary Cumulative Distribution Function denotes the probability that
the variable takes a value greater than x.
In our case, the CSFD indicates the probability that the energy of an icequake will take
a value greater than a given value. 

In order to study the temporal variation of the CSFD, we used a moving window
of 200 events with a 20-event shift
between successive windows. We analyzed the event size distribution
corresponding to each window (Fig. \ref{fig3}, top, shows three typical windows).
The exponent $\beta$ (Fig. \ref{fig3}, bottom) was estimated using the Maximum-Likelihood fitting
method with goodness-of-fit tests based on the Smirnov test (often improperly known as the Kolmogorov-Smirnov test) (see for instance \citet{Clauset&al2009}
for a review on practical issues and empirical analyses). For each
exponent, we used a goodness-of-fit test, which generates a p-value that
quantifies the plausibility of the power law hypothesis.
The purpose of this test is to sample many synthetic data sets from a true power-law
distribution and to measure how far they fluctuate away from the power law form,
and to compare the results with similar measurements of the empirical
data. The quantification of the distance between two distributions was made
using the Kolmogorov-Smirnov statistics.
The p-value was defined as being the fraction of the synthetic distances that were
larger than the empirical distance.
If p was not too small, the difference between the empirical and the synthetic data could be attributed to
statistical fluctuations alone; if $p<0.1; 0.05$, the fit was poor and the model was
not appropriate at the $90\%; 95\%$ confidence level. 
By applying the p-test on our data, we obtained p-values
greater than 0.2, indicated in Fig. \ref{fig3} by empty symbols . 
This corresponds to the time windows for which we cannot reject the hypothesis
that the CSFD was indeed generated from a power law distribution.

Three different behaviors were observed in succession:

\noindent(i) For the windows located near the beginning of our
measurements (up to $t=14\,~\rm d$), the CSFD was well described by a power-law distribution over at least 3 orders of magnitude (see upper left
panel of Fig. \ref{fig3}), indicating
a scale invariance of the acoustic emissions.

\noindent(ii) From $t=14\,~\rm d$ to $t=8 \,~\rm d$, the exponent
$\beta$ exhibited a rapid shift, 
suggesting a change in
behavior of the damage process developing in the ice mass.
As the p-values were low in this period, the power-law behavior was not a
plausible fit to the data.

\noindent(iii) For the time windows near the end of our observation period
(after $t=8 \,\rm d$), the CSFD recovered a power-law behavior, with a
high $\beta$-value, with a shoulder at the tail of the distribution
(upper right panel of Fig. \ref{fig3}).

\subsection{Waiting time distribution and accelerating rate of icequakes}


\begin{figure}
\vspace*{2mm}
\begin{center}
\includegraphics[width=0.7\textwidth]{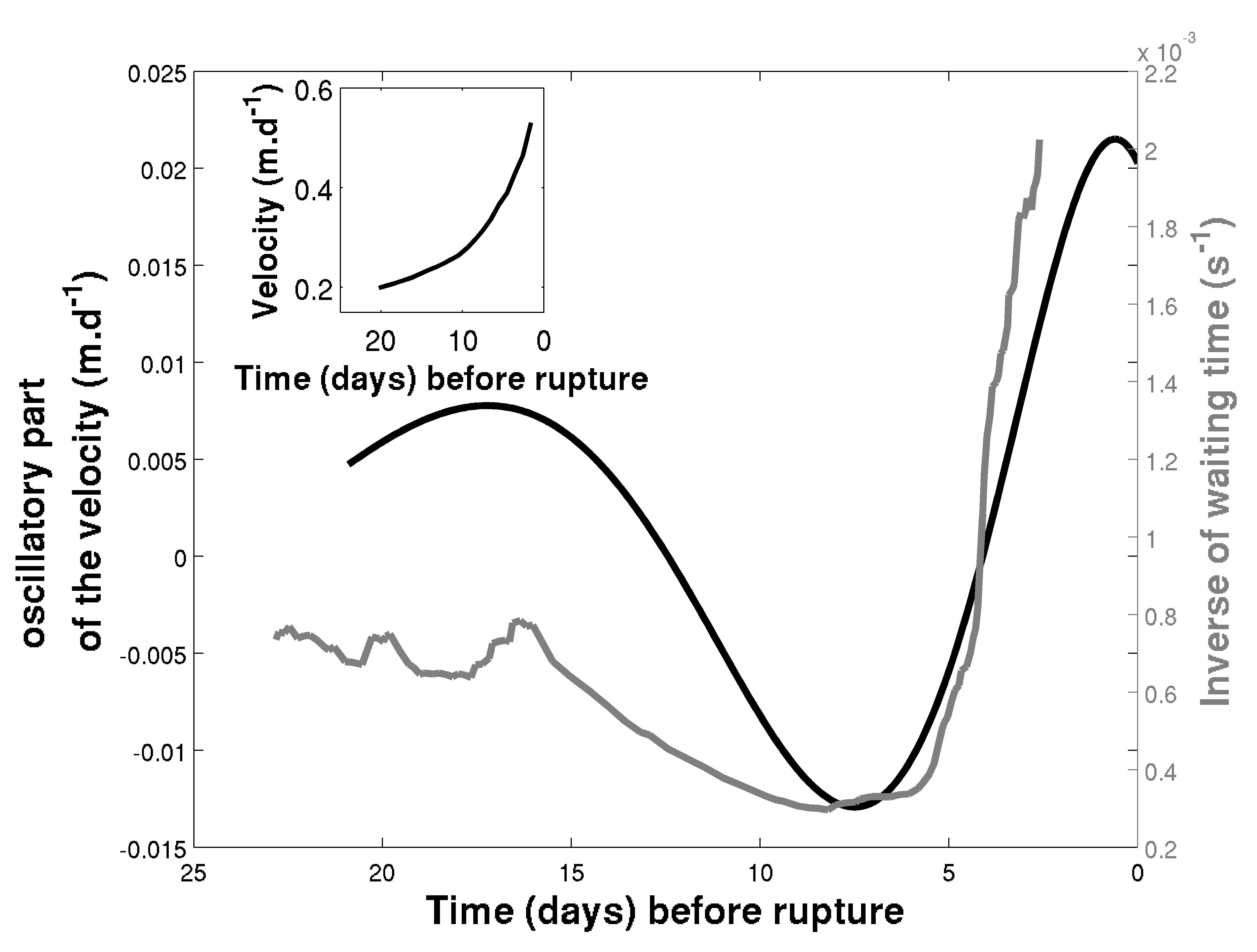}
\end{center}
\caption{\label{comp} Plot of the inverse of the waiting time 
between successive icequakes (noisy grey curve) and of the
oscillatory part of the evolution of the surface velocity (smooth dark oscillatory curve).
Inset: surface velocity as a function of time up to the first break-off.}
\end{figure}

The time evolution of the rate of icequakes is well-captured by 
the inverse of the mean time lag between two consecutive
icequakes. Fig. \ref{comp} shows this inverse mean time lag (which can be associated with a mean frequency of icequake events, i.e. the seismic activity) 
in a moving window containing 100 events as a function of the time of the last point of
the window. 
One can clearly observe a general acceleration of the icequake
activity approximately one week before the break-off of the glacier.
The size of the moving window was set as small as possible to enable an accurate practical detection of the acceleration of the icequake activity.

\begin{figure}
\vspace*{2mm}
\begin{center}
\includegraphics[width=0.7\textwidth]{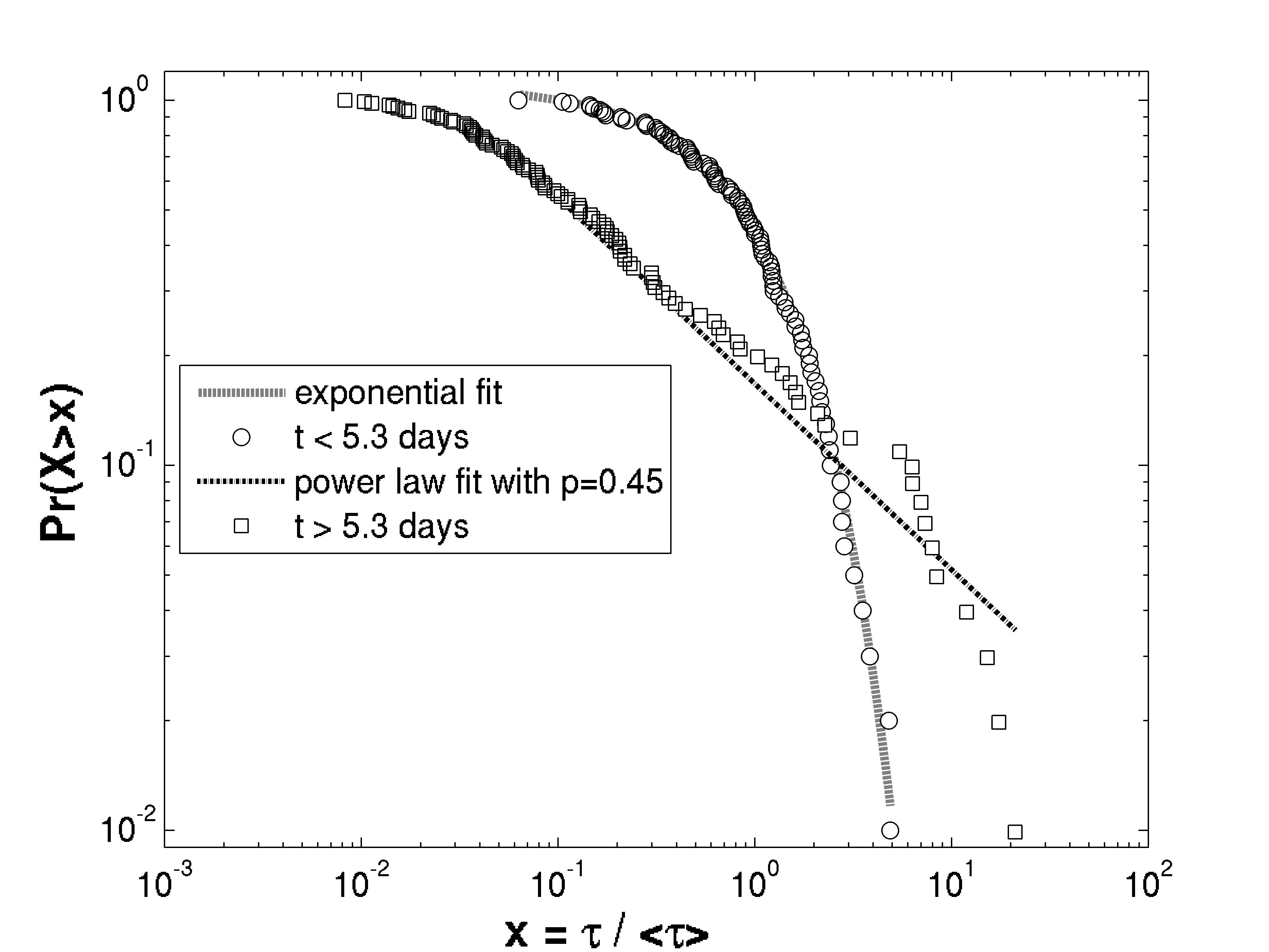}
\end{center}
\caption{\label{distiwt} Complementary waiting time distribution ($Pr(X>x)$) for the 100 events before and after the
transition ($\simeq$ 5.2 days) between stable and unstable regimes. $\tau$
is the waiting time between two icequakes, $\langle \tau \rangle$ is the mean of all the waiting times considered. The data  for $t \leq 5.2$ days can be well fitted by 
the exponential function $p(x) \sim a \cdot \exp(bx)$ with $a = 110$ and $b = -0.93$. 
For $ t \geq 5.2$ days, the distribution of waiting times was compatible with a power law
$p(x) \sim x^{-\alpha}$ for $x > x_{min}$ with $\alpha$=1.5
and $x_{min}=0.058$. }
\end{figure}

We performed the same statistical analysis as for icequake energy (see Section
\ref{energy}).
Fig. \ref{distiwt}, which shows the complementary cumulative size-frequency distribution of waiting time between two icequakes, 
exhibits a change in the waiting time
distribution as the global rupture is approached.
It appears that this distribution is initially well described by a power law distribution, indicating a temporal
correlation between the icequakes. A few days
before the glacier break-off, the waiting times distribution shifted to an exponential
distribution, indicating a loss of temporal correlation between the
icequakes. 




\subsection{Surface displacements}

Thanks to very accurate displacement measurements ($< 1 \rm cm$),
\citet{Faillettaz&al2008} were able to demonstrate the existence of log-periodic
oscillations superimposed on the power law acceleration. 
Log-periodicity, that is periodicity in the logarithm of the time-to-rupture $t_c-t$ 
(where the rupture occurs at time $t_c$),
is the empirical signature of the symmetry of discrete scale invariance, which means that 
the observable is self-similar to itself only under integer powers of a fundamental
scaling ratio $\lambda$ of time scales (see review and details in Sornette (1998)). 

This log-periodic behavior can be shown by the following 
equation describing the surface displacement as a function of time:
\begin{equation}
\label{logper}
s(t)=s_0+a(t_c-t)^{m}\Big[1+C\sin\left(2\pi\frac{\ln(t_c-t)}{\ln(\lambda)}+D\right)\Big]~.
\end{equation}
Here, $s_0$ is a constant, $t_c$ is the critical time at which the 
global collapse is expected, $m<1$ is the power
law exponent quantifying the acceleration, $a$ is a constant, $C$ is the relative amplitude
of the oscillations with respect to the overall power law acceleration, $\lambda$ is the
so-called ``scaling ratio'' associated with the log-periodicity of expression (\ref{logper})
and $D$ is the phase of the log-periodic oscillation.


\section{Interpretations and discussions}

\subsection{Size-frequency distributions of icequake energy}
As described in Section \ref{energy}, three different regimes can be identified
during the maturation of the rupture event:

\begin{itemize}
\item[(i) ]  The size-frequency distribution of icequake energy exhibits a
 power-law behavior, indicating
a scale invariance of the acoustic emissions, similar to that
characterizing earthquakes. For earthquakes, the corresponding
Gutenberg-Richter law describes one of the most ubiquitous
statistical regularities observed (see for instance Pisarenko and Sornette (2003) and
references therein). It reads
\begin{equation}
\label{GR}
N(>E) \sim E^{-\beta}\;,
\end{equation}
where $N(>E)$ is the number of events with an energy greater than $E$ and $\beta$ is 
the Gutenberg-Richter exponent found empirically close to $2/3$ for
shallow earthquakes (depths $< 70$ km) in subduction and transform fault
zones. 
Up to $t~=~14$~d, the exponent $\beta$ is found to be compatible with the
earthquake value $\beta =0.65 \pm 0.1$.

This behavior can be associated with a stable critical regime, in accordance
with the critical behavior of rupture found in sufficiently heterogeneous
media \citep{Johansen&Sornette2000}, similar to critical phase transitions (from diffuse damage to crack
nucleation).
As this non-equilibrium system exhibits a kind of dynamic criticality
without tuning any control parameter, this regime could be associated with a self-organized
critical (SOC) behavior. We refer to Jensen (1998), Turcotte (1999) and Sornette (2006)
for introductions and detailed presentations of the concepts and models of self-organized criticality.
The first stable regime fulfills the characteristics of SOC:
(a) power-law distributions, such as those observed in this regime, (b)
a constant driving stress, (c) a threshold dynamics and (d) a very
large number of interacting local entities (here, the micro cracks).
In other words, in a such regime, the glacier has time to adapt to the new state induced by the rupture
maturation process.

\item[(ii) ] In this transitional regime, the glacier can no longer adapt itself to the
changes induced by rupture maturation, and the CSFD of icequakes is no
longer a power law. A main feature of the extreme tail is the large ``characteristic events'',
that can be interpreted as the nucleation of the incipient rupture.

\item[(iii) ] The size-frequency distribution of icequake energy exhibts a
 power-law behaviour with lower b-value and an appearance of characteristic events.
\citet{Pisarenko&Sornette2003} have associated a change in the exponent $\beta$
with a change in the rupture process.
They proposed the following explanation of these two regimes:
First, large exponents $\beta$ are found in the distribution of acoustic emission energies
recorded for heterogeneous materials brought
to rupture, for which damage occurs mainly in the form of weak shear zones and open cracks.
In other words, large exponents $\beta$ are meanly an indication of an open crack mode of damage.
Second, when damage develops in the form of `` dislocations'' 
or mode II cracks, with slip mode of failure and with healing, 
the exponent $\beta$ is found to be smaller than $1$. 

This suggests that  the low value of $\beta \simeq 2/3$ found up to two weeks
before rupture  is associated with a
stable, slow and diffuse ``dislocation-like'' damage process. 
In the subsequent days,  the increase in the exponent $\beta$, together with the developing shoulder 
at high icequake energies, can then be interpreted as revealing a transition to a mode of damage
controlled more and more by crack openings and their coalescence prior to the incipient rupture. 

The tail of the distribution develops a strong shoulder, indicating a change in the 
damage evolution process. The clear deficit of icequakes with low energies and the
excess of large ``characteristic'' events (events with high energy) is fully compatible with the 
evolution of the second regime dominated by crack-like events which, by
their proliferation and fusion, progressively nucleate the formation of the run-away macro-crack
responsible for the final avalanche associated with a rather clean
crack-like rupture, as shown in Fig. \ref{general}. 
Moreover, as coupling between cracks increases, and heterogeneity decreases, 
the power law develops a shoulder, corresponding to 
a different regime of global rupture \citep{Sornette2009}.
\end{itemize}

\subsection{Seismic activity}

Acoustic emissions were also observed experimentally in
heterogeneous materials brought to rupture and were found to exhibit clear acceleration, in
agreement with a power-law divergence expected from the critical point theory
\citep{Johansen&Sornette2000}. In a nutshell, critical point theory views
the global rupture of a system as the result of a progressive organization 
process of defects. The global rupture corresponds to a 'bifurcation' (also known as 'phase transition',
'catastrophe' or 'tipping point'), resulting from the collective organization of defects
that interact to prepare the global transition, i.e., the rupture
\citep{SorSor1990,SorVan1990,Anifranietal1995,Andersenetal1997,SorAnd1998,SamSornette2002}.
The concept of critical rupture suggests that global ruptures are preceded
by specific precursors that make them predictable in a probabilistic way
\citep{Anifranietal1995,Sornette2002}.

Our results also support this picture as they clearly indicate a general acceleration in icequake
activity approximately one week before the break-off event.
Recent studies in material sciences also show a clear transition to the
tertiary creep regime with an acceleration in the rate of acoustic emissions
\citep[and references therein]{Nechad&al2005a,Nechad&al2005b}.

\subsection{Comparison between icequake activity and surface displacements}


\citet{Faillettaz&al2008} demonstrated a general power-law acceleration of the surface
displacement before the break-off of this hanging glacier. Moreover, this acceleration was accompanied by
oscillations which increased logarithmically in frequency
as the time of failure approached (referred to as log-periodic
oscillations).
The origin of such oscillations is not yet fully understood,
but dynamic crack interactions were hypothesized as a possible mechanism leading to such
log-periodic oscillations \citep{Saleuretal1996,Huang&al1997,IdeSor2002}.

The discrete hierarchy of the organization
of damage revealed by the log-periodic oscillations \citep{Faillettaz&al2008}
should somehow influence
the seismic activity of the glacier during the maturation of the rupture process.
Fig. \ref{comp} shows the comparison between the oscillatory part of the
velocity (time derivative of the surface displacements which follow a
log-periodic power law) and the inverse of the waiting time between successive
icequakes.

It appears that these two metrics exhibit a significant correlation (with $r^2 = 0.65$).
This result means that the seismic activity is not correlated with the global
power-law acceleration of the glacier during the rupture maturation but rather with
its superimposed oscillations. It is more the jerky motions around the overall accelerations
that are responsible for detectable seismic activity \citep{JohSor99}, corresponding to crack growth and coalescence events
within the glacier.




\subsection{Waiting time distribution}

The waiting time distribution was seen to be first power law distributed, and
then it shifted to an exponential distribution.
This result is in accordance with the previous physical
interpretation of crack coalescence.
We tentatively attribute this change of the distribution of waiting times from power-law to exponential
to the transition from diffuse damage to damage
organizing in the form of clusters.

The random activation of different damage clusters when approaching the global failure causes a
transient loss of the temporal correlation of the individual fracture
events \citep{Kuksenko&al2005}. 
This effect confirms the existence of a hierarchical structure of the fracture process
in the glacier.

In contrast to the difficulties in ascertaining the existence of an 
accelerating moment release upon the approach to a large earthquake
\citep{Hardebeck&al2008}, the four metrics that we applied to our glacier data
support the existence in this case of an accelerated deformation and damage process
occurring in the last few days before the incipient break-off.

\subsection{Evolution of failure processes leading to the break-off event}

It is now possible to draw the following picture of the evolution of the failure processes leading to the break-off event.

\begin{enumerate}
\item[(i)] {\bf From the beginning of our measurements to 15 days before the break-off event}:
In this regime, seismic activity is more or less constant, the size-frequency distribution of icequake energies 
is described by a power-law behavior and the waiting time
distribution is also a power law.
This behavior could be related to a self-organizing regime, where diffuse damage
accumulates within the glacier, with a proliferation of dislocation-like defects.
In other words, the glacier has time to adapt to the deformation and damage maturation process.

\item[(ii)] {\bf From 15 days to 5 days before the break-off event}:
seismic activity is slightly decaying, the size-frequency
distribution is no longer power-law, the waiting time distribution is power law
and the glacier is decelerating relative to the previous phase. 
This is a transitional regime: the damage process goes on, micro-cracks grow and
start merging in a homogeneous
way. Log-periodic oscillations appear and reveal the hierarchical structure of the fracture process under development.
In such a regime, screening and locking effects are likely to appear, possibly explaining the slight decay of
icequake activity and the relative slowdown of surface velocity. 


\item[(iii)] {\bf From 5 days before the final break-off event}: 
Seismic activity drastically increases, the size-frequency
distribution of icequake energy recovers its power-law behavior but with
a different exponent, together with the appearance of some large characteristic events
leading to a strong shoulder in the distribution. The waiting
time distribution exhibits a shift from a power law to an exponential distribution.
The system enters into a catastrophic regime where damage clusters are randomly activated. 
Damage clusters interact and merge with a preferential direction (i.e. preparing the
final rupture pattern), in contrast to the previous regime. 
The largest scale of the hierarchical structure of the fracture process is activated
(resulting in characteristic events).


\end{enumerate}

\section{Summary and perspectives}

We have presented and studied a unique data set of icequakes recorded
in the immediate vicinity of a hanging glacier over 25 days prior to and up to its rupture.
While seismic measurement records unfortunately ceased 3 days before the first
break-off and 10 days before the larger subsequent one, we
nevertheless were able to obtain a coherent quantitative picture of the 
damage process developing before the impending glacier collapse.
Our main results include first the demonstration of
a clear increase in the icequake activity within the glacier (measured as the
inverse of the waiting time between successive icequakes) 
starting approximately $6$ days before the first break-off.
Secondly, we found a two-step evolution of the
size-frequency distribution of icequake energies, characterizing
a first transition to a crack-like dominated damage followed by a second
transition in which large characteristic cracks prepared the nucleation
of the run-away rupture. Thirdly, we documented
the acceleration of the rate of icequakes combined with a change in the
distribution of waiting times between icequakes, which is typical for the hierarchical
cascade of rupture instabilities found in earlier reports on the acoustic
emissions associated with the failure of heterogeneous materials.
Moreover, a clear correlation between the seismic activity and the oscillatory
part of the surface velocity evolution prior to the first break off event was
also documented.
By combining the analysis of surface motion with log-periodic oscillations and icequake activity during the rupture maturation process, prediction of
the final break-off time was improved significantly.
This also provides new insights into the physical mechanisms of the rupture in heterogeneous
materials.

Provided that technical solutions can be found to ensure continuous
icequake recordings in the difficult high-altitude mountain 
conditions, our results clear the path for real-time diagnostics
of impending glacier failure. The next steps towards this goal include
(a) developing an automatic seismic data processing in real time
(which includes the automatic detection of icequakes and the determination
of their energy), (b) processing these data with the statistical tools
developed here, and (c) performing systematic reliability tests to access
the rate of false alarms (false positives or errors of type I) versus missed 
events (false negatives or errors of type II). Step (c) is necessary
for an informed cost-benefit analysis of the societal and economic
impacts of the proposed real-time forecast methodology.

{\bf Acknowledgements}:
The Institute of Geophysics at the ETH Z\"urich is gratefully acknowledged for
allowing us the use of their instruments.
Thanks are extended to Dr. Fabian Walter for fruitful
discussions. Dr. Paul Winberry, as well as an anonymous referee and Ted Scambos contributed
substantially to the improvements made in the manuscript. We are also grateful to C. Wuilloud (natural hazards, Valais) for
logistic and financial support.


\end{document}